\documentclass
[notitlepage,12pt,groupedaddress,secnumarabic,prd,twocolumn,showpacs,showkeys]{revtex4}%
\usepackage{amssymb}
\usepackage{amsfonts}
\usepackage{amsmath}
\usepackage{latexsym}
\usepackage{graphicx}
\usepackage{appendix}%
\begin{document}
\title{Short note on the stability of a dilatonic wall\\ }
\author{J.R. Morris}
\affiliation{Physics Dept., Indiana University Northwest, 3400 Broadway, Gary, Indiana,
46408, USA\ \ }

\begin{abstract}
A nontopological soliton solution of dilaton-Maxwell theory describes a domain
wall-like solution which confines magnetic flux in its core [G.W. Gibbons and
C.G. Wells, Class. Quant. Grav. 11, 2499 (1994)]. Since the solution is not
stabilized by a nontrivial topology of the vacuum manifold, it is interesting
to see if the static solution is stable againt small fluctuations. We consider
the stability of the solution in response to small fluctuations in the scalar
and magnetic fields. It is determined that the ansatz solution does indeed
exhibit stability.

\end{abstract}

\maketitle

\section{Introduction: Dilaton-Maxwell theory}

\ \ Topological solitons are stabilized by a nontrivial topology of the vacuum
manifold of a scalar field, but nontopological solitons lacking a nontrivial
topology of a vacuum manifold can also exist. Sometimes nontopological
solitons can be stabilized by fermions (see, for example, \cite{DS1}%
-\cite{FB}) or bosons (see, for example, \cite{nts}-\cite{bub}) that interact
with or become trapped within the solitons. However, presently we focus
attention on a peculiar and interesting nontopological soliton \cite{GW} that
arises in dilaton-Maxwell theory, without any interactions with fermions or
other scalar bosons, and attempt to determine the stability of the soliton.
\smallskip\smallskip

\ \ We consider here the form of dilaton-Maxwell theory in a flat spacetime
($g_{\mu\nu}=\eta_{\mu\nu}$) described by an action \cite{GW}%
\begin{equation}
S=\int d^{4}x\left(  \frac{1}{2}\partial_{\mu}\phi\partial^{\mu}\phi-\frac
{1}{4}e^{-2\tilde{\kappa}\phi}F_{\mu\nu}F^{\mu\nu}\right)  \label{1}%
\end{equation}

where $\tilde{\kappa}$ is an arbitrary real constant with dimension of inverse
mass, so that $\tilde{\kappa}\phi$ is dimensionless, and $F_{\mu\nu}%
=\partial_{\mu}A_{\nu}-\partial_{\nu}A_{\mu}$ is the electromagnetic field
tensor. (We use a metric with negative signature $(+,-,-,-)$.) We have
\begin{equation}
F_{\mu\nu}=\left(
\begin{array}
[c]{cccc}%
0 & E_{x} & E_{y} & E_{z}\\
-E_{x} & 0 & -B_{z} & B_{y}\\
-E_{y} & B_{z} & 0 & -B_{x}\\
-E_{z} & -B_{y} & B_{x} & 0
\end{array}
\right)  \label{2}%
\end{equation}

The equations of motion following from (\ref{1}), along with the Bianchi
identity, are
\begin{subequations}
\label{3}%
\begin{gather}
\square\phi+\frac{1}{2}\tilde{\kappa}e^{-2\tilde{\kappa}\phi}F_{\mu\nu}%
F^{\mu\nu}=0\label{3a}\\
\nabla_{\mu}\left(  e^{-2\tilde{\kappa}\phi}F^{\mu\nu}\right)
=0,\ \ \ \ \ \nabla_{\mu}\ \tilde{F}^{\mu\nu}=0 \label{3b}%
\end{gather}

where $\tilde{F}_{\mu\nu}=\frac{1}{2}\epsilon_{\mu\nu\rho\sigma}F^{\rho\sigma
}$ is the dual electromagnetic field tensor. The equations of (\ref{3b}) can
be rewritten in terms of electric and magnetic fields as%
\end{subequations}
\begin{align}
\nabla\cdot\mathbf{D}  &  =0,\ \ \nabla\times\mathbf{H}-\mathbf{\dot{D}%
}=0,\nonumber\\
\nabla\cdot\mathbf{B}  &  =0,\ \ \nabla\times\mathbf{E}+\mathbf{\dot{B}}=0
\label{4}%
\end{align}

with $\mathbf{D}=\epsilon\mathbf{E}$ and $\mathbf{B}=\mu\mathbf{H}$. Here the
effective dielectric function $\epsilon$ and effective permeability function
$\mu$ are given by%
\begin{equation}
\mu=\epsilon^{-1}=e^{2\tilde{\kappa}\phi} \label{5}%
\end{equation}

so that the index of refraction is unity. Let us rewrite the equation of
motion (\ref{3a}) for the real scalar field $\phi$ in the form%
\begin{align}
\nabla^{2}\phi-\partial_{t}^{2}\phi &  =-\tilde{\kappa}e^{-2\tilde{\kappa}%
\phi}(\mathbf{B}^{2}-\mathbf{E}^{2})\nonumber\\
&  =-\tilde{\kappa}e^{2\tilde{\kappa}\phi}(\mathbf{H}^{2}-\mathbf{D}^{2})
\label{6}%
\end{align}

\bigskip

\ \ Gibbons and Wells \cite{GW} have found a couple of interesting
nontopological solitonic solutions to these equations of motion, one
describing a type of cosmic string which confines magnetic field $\mathbf{B}$
and magnetic flux $\Phi_{\text{mag}}$ in its core, and another wall-like
solution which also confines the magnetic field $\mathbf{B}$ and magnetic flux
within its core. Here we focus upon the solitonic Gibbons-Wells wall solution
to (\ref{4}) and (\ref{6}), which is easier to study, and consider its
stability against decay.

\section{Static domain wall ansatz solution}

\ \ Gibbons and Wells discovered an interesting magnetic wall solution to
(\ref{4}) and (\ref{6}) using an ansatz where $\mathbf{E}=\mathbf{D}=0$ and
$\mathbf{H}=(0,0,\mathcal{H})=$ const. and $\mathbf{B}=(0,0,\mathcal{B}%
)=\mu\mathbf{H}$. Furthermore, it is assumed that the scalar field $\phi$
takes the form $\phi=\Phi(x,y)$. The equation of motion for $\Phi$ is then
given by%
\begin{equation}
(\partial_{x}^{2}+\partial_{y}^{2})\Phi=-\tilde{\kappa}\mathcal{H}%
^{2}e^{2\tilde{\kappa}\Phi} \label{7}%
\end{equation}

This is recognized as the 2D Euclidean Liouville equation whose solution is
given by \cite{GW},\cite{Liouville},\cite{Crowdy}%
\begin{equation}
\mu(\zeta)=e^{2\tilde{\kappa}\Phi(\zeta)}=\frac{4}{\tilde{\kappa}%
^{2}\mathcal{H}^{2}}\frac{\left\vert f^{\prime}(\zeta)\right\vert ^{2}%
}{\left(  1+\left\vert f(\zeta)\right\vert ^{2}\right)  ^{2}}\label{8}%
\end{equation}

where $\zeta=x+iy$ and $f(\zeta)$ is a holomorphic function of $\zeta$. For
the static wall ansatz solution, $f(\zeta)$ is chosen to take the form
$f(\zeta)=\exp(M\zeta)$. Then (\ref{8}) yields the solution%
\begin{align}
\mu(x)  &  =e^{2\tilde{\kappa}\Phi(x)}=\left(  \frac{M}{\tilde{\kappa
}\mathcal{H}}\right)  ^{2}\frac{1}{\cosh^{2}(Mx)}\nonumber\\
&  =\left(  \frac{M}{\tilde{\kappa}\mathcal{H}}\right)  ^{2}\text{sech}%
^{2}(\bar{x}),\ \ \ \ \bar{x}\equiv Mx \label{9}%
\end{align}

The parameter $M$ has dimension of mass so that the coordinate $\bar{x}=Mx$ is
dimensionless, as is the factor $(M/\tilde{\kappa}\mathcal{H})$. Note that
this ansatz solution depends only on $x$, and not on $y$.

\bigskip

\ \ Using (\ref{1}) along with the ansatz, we write the lagrangian of this
ansatz system as%
\begin{align}
\mathcal{L}  &  =\frac{1}{2}\partial^{\mu}\phi\partial_{\mu}\phi-\frac{1}%
{4}e^{-2\tilde{\kappa}\phi}F_{\mu\nu}F^{\mu\nu}\nonumber\\
&  =\frac{1}{2}\partial^{\mu}\Phi\partial_{\mu}\Phi+\frac{1}{2}\mathcal{H}%
^{2}e^{2\tilde{\kappa}\Phi} \label{10}%
\end{align}

and identify the scalar potential as $V(\Phi)=-\frac{1}{2}\mathcal{H}%
^{2}e^{2\tilde{\kappa}\Phi}$. We see that $\mu(x)\rightarrow0$ and
$\mathbf{B}(x)\rightarrow0$ as $|x|\rightarrow\infty$ for the wall solution of
(\ref{9}). The magnetic $\mathbf{B}$ field is described by $\mathcal{B}%
(x)=\mu(x)\mathcal{H}\propto$ sech$^{2}(Mx)$ and the magnetic flux per unit
length of the wall is%
\begin{equation}
\frac{\Phi_{\text{mag}}}{L_{y}}=\frac{1}{L_{y}}\int_{-\infty}^{\infty}\int
_{0}^{L_{y}}\mathcal{B}(x)dxdy=\frac{2M}{\tilde{\kappa}^{2}\mathcal{H}}
\label{11}%
\end{equation}

\section{Stability}

\ \ We now consider the stability of the Gibbons-Wells magnetic wall solution
of (\ref{9}) in response to small fluctuations of the scalar field $\phi$ (and
therefore fluctuations in $\mathbf{B}$) about its ansatz value $\Phi(x)$. In
particular, we want to investigate the possibility of any uncontrolled growth
or decay of the ansatz solution. We write $\phi(x,y,z,t)=\Phi(x)+\delta
\phi(x,y,z,t)$ and insert this into the equation of motion (\ref{6}) while
maintaining $\mathbf{D}=0$ and $\mathbf{B}\rightarrow\mathbf{B}^{\prime
}(x,y,z,t)=\mu^{\prime}(x,y,z,t)\mathbf{H}$ with $\mathbf{H}=(0,0,\mathcal{H}%
)=$ const and $\mu^{\prime}=\exp(\Phi(x)+\delta\phi)$. Keeping only terms
linear in $\delta\phi$ and using (\ref{7}) we have%
\begin{equation}
\nabla^{2}\delta\phi-\partial_{t}^{2}\delta\phi=-2(\tilde{\kappa}%
\mathcal{H})^{2}e^{2\tilde{\kappa}\Phi(x)}\delta\phi\label{12}%
\end{equation}

where we have used ($e^{2\tilde{\kappa}\delta\phi}-1)=2\tilde{\kappa}%
\delta\phi$. The time dependence of $\delta\phi$ is taken to be given by
$\delta\phi(x,y,z,t)=\Psi(x,y,z)\sin\omega t$ and solutions of (\ref{12}) with
$\omega^{2}<0$ will signal an instability of the solution $\Phi(x)$, while
solutions with $\omega^{2}\geq0$ signal stability against dissipation. We
therefore arrive at the equation for $\Psi(\mathbf{x})$ given by%
\begin{equation}
-\nabla^{2}\Psi-\frac{2M^{2}}{\cosh^{2}(\bar{x})}\Psi=\omega^{2}\Psi\label{13}%
\end{equation}

It is convenient to define the dimensionless variables $\bar{x}^{\mu}=Mx^{\mu
}=(\bar{t},\bar{x},\bar{y},\bar{z})$ and $\bar{\omega}=\omega/M$ and rewrite
(\ref{13}) as%
\begin{equation}
-\bar{\nabla}^{2}\Psi(\mathbf{\bar{x}})-2\text{sech}^{2}(\bar{x}%
)\Psi(\mathbf{\bar{x}})=\bar{\omega}^{2}\Psi(\mathbf{\bar{x}}) \label{14}%
\end{equation}

where $\bar{\nabla}^{2}$ is the Laplacian in terms of coordinates $\bar{x}$,
$\bar{y}$, and $\bar{z}$.

\bigskip

\ \ We next perform a separation of variables by writing $\Psi(\mathbf{\bar
{x}})=\psi(\bar{x})f(\bar{y})g(\bar{z})$ so that (\ref{14}) leads to%
\begin{equation}
-\frac{1}{\psi}\partial_{\bar{x}}^{2}\psi-\frac{1}{f}\partial_{\bar{y}^{2}%
}f-\frac{1}{g}\partial_{\bar{z}^{2}}g+U(\bar{x})=\bar{\omega}^{2} \label{15}%
\end{equation}

with
\begin{equation}
U(\bar{x})=-2\text{sech}^{2}(\bar{x}) \label{15a}%
\end{equation}

Setting $-\frac{1}{f}\partial_{\bar{y}}^{2}f=\bar{k}_{y}^{2}$ and $-\frac
{1}{g}\partial_{\bar{z}}^{2}g=\bar{k}_{z}^{2}$ we have a solution of the form
$\Psi(\bar{x},\bar{y},\bar{z})\sim\psi(\bar{x})\sin(\bar{k}_{y}\bar{y}%
+\delta_{y})\sin(\bar{k}_{z}\bar{z}+\delta_{z})$ where $\bar{k}_{y}^{2}%
+\bar{k}_{z}^{2}$ must be real valued. Now (\ref{15}) reduces to%
\begin{equation}
-\partial_{\bar{x}}^{2}\psi(\bar{x})+U(\bar{x})\psi(\bar{x})=[\omega^{2}%
-(\bar{k}_{y}^{2}+\bar{k}_{z}^{2})]\psi(\bar{x})\label{16}%
\end{equation}

We now appeal to the Maxwell equations of (\ref{4}) with $\mathbf{E}%
=\mathbf{D}=0$ with replacements $\mathbf{B}\rightarrow\mathbf{B}^{\prime}%
=\mu^{\prime}\mathbf{H}=\mu^{\prime}(0,0,\mathcal{H})$, where%
\begin{equation}
\mu^{\prime}=e^{2\tilde{\kappa}\phi}=e^{2\tilde{\kappa}\Phi(x)}e^{2\tilde
{\kappa}\delta\phi}=\mu(1+2\tilde{\kappa}\delta\phi) \label{17}%
\end{equation}
with $\mu$ given by (\ref{9}). In order to satisfy the Maxwell equations we
require $\nabla\cdot\mathbf{B}^{\prime}=0$ and $\mathbf{\dot{B}}^{\prime}=0$,
which, in turn, imply that $\partial_{z}\mu^{\prime}=\partial_{z}\delta\phi=0$
and $\partial_{t}\mu^{\prime}=\partial_{t}\delta\phi=0$, requiring the
conditions $\bar{k}_{z}=0$ and $\omega=0$. The result $\omega=0$ signals an
absence of instability of the ansatz solution, but we must check that
$\Psi(\mathbf{x})$ satisfies our consistency condition that $\delta\phi$
remains small for all $\mathbf{x}$.

\bigskip

\ \ Equation (\ref{16}) has now reduced to%
\begin{equation}
-\partial_{\bar{x}}^{2}\psi(\bar{x})+U(\bar{x})\psi(\bar{x})=-\bar{k}_{y}%
^{2}\psi(\bar{x})\equiv2E\psi(\bar{x}) \label{18}%
\end{equation}

Since $\psi(\bar{x})$ and $U(\bar{x})$ are real we require $\bar{k}_{y}^{2}$
to be real. In order that $f(y)$ be continuous and asymptotically finite, we
further require that $\bar{k}_{y}$ be real. Then we have $-\bar{k}_{y}^{2}%
\leq0$. Let us set $-\bar{k}_{y}^{2}=2E$, where $E$ is the \textquotedblleft
energy\textquotedblright\ parameter of the Schr\"{o}dinger equation
(\ref{18}). In fact, (\ref{18}) can be recognized as the Schr\"{o}dinger
equation with a P\"{o}schl-Teller potential%
\begin{equation}
U(\bar{x})=-\lambda(\lambda+1)\text{sech}^{2}(\bar{x}) \label{19}%
\end{equation}

A comparison of (\ref{15a}) and (\ref{19}) requires the setting $\lambda=1$.
The solution to (\ref{18}) for $\lambda=1$ is the bound state with
\textquotedblleft energy\textquotedblright\ $E=-1/2$ i.e., $\bar{k}_{y}=\pm1$,
and $\psi(\bar{x})=A$sech$(\bar{x})$ with $A$ being a real constant. We
therefore have a solution for $\delta\phi$ which can be kept small for all
$\mathbf{x}$ and has $\omega=0$, satisfying the conditions for stability of
the ansatz solution $\Phi(x)$ with respect to small fluctuations in the scalar
field $\phi$ and magnetic field $\mathbf{B}$. In addition to the static
solution $\delta\phi(x,y)$, there also exists a class of generalized magnetic
wall solutions which can support arbitrarily large, nondissipating traveling
waves \cite{JM06} having the form $\phi(x)\rightarrow\phi(X)$, where
$X=x-f(t\pm z)$ with $\square f=0$ and $\partial_{\mu}f\partial^{\mu}f=0$.
(These types of \textquotedblleft wiggly\textquotedblright\ solutions are also
supported by topological domain walls and cosmic strings \cite{Vach}.)

\bigskip

\ \ In addition, we expect the wall to be stable against bending and
spontaneous bubble formation. Our reasoning is that the lagrangian for the
field $\phi$ is $\mathcal{L}=\frac{1}{2}\partial^{\mu}\phi\partial_{\mu}%
\phi-V(\phi)$, where $V(\phi)=-\frac{1}{2}\mathcal{H}^{2}e^{2\tilde{\kappa
}\phi}$ [see (\ref{10})] and the scalar field stress-energy is $T_{\mu\nu
}=\partial_{\mu}\phi\partial_{\nu}\phi-\eta_{\mu\nu}\left[  \frac{1}%
{2}\partial^{\alpha}\phi\partial_{\alpha}\phi-V\right]  $. This gives an
energy density for a static wall of $T_{00}=\frac{1}{2}|\nabla\phi
|^{2}+V=\frac{1}{2}|\nabla\phi|^{2}-\frac{1}{2}\mathcal{H}^{2}e^{2\tilde
{\kappa}\phi}$. Both $\phi$ and $V$ are symmetric functions of $x$, so that a
slice of wall of thickness $|\delta x|$ at a distance $|x|$ from the center
contributes to an amount of surface tension (energy per unit area) of
magnitude $|\delta\mu|=|\int_{x}^{x+\delta x}T_{00}dx|$, which should be the
same for two slices at positions $\pm|x|$. Since the surface tension is the
same on both sides, i.e., there is no biasing in the energy density, we expect
stability against bending, although the wall can support traveling waves
\cite{JM06}.

\bigskip

\ \ From the ansatz solution $\Phi(x)$ given by (\ref{9}), we compute the
scalar field and magnetic energy densities to be
\begin{subequations}
\label{20}%
\begin{align}
T_{00}^{(\Phi)}  &  =\frac{M^{2}}{2\tilde{\kappa}^{2}}\left(  \tanh^{2}\bar
{x}-\text{sech}^{2}\bar{x}\right) \label{20a}\\
T_{00}^{(\mathbf{B})}  &  =\frac{1}{2}\mathbf{H\cdot B}=\frac{1}{2}%
\mu(x)\mathcal{H}^{2}\nonumber\\
&  =\frac{M^{2}}{2\tilde{\kappa}^{2}}\text{sech}^{2}\bar{x} \label{20b}%
\end{align}

so that the total energy density is simply%
\end{subequations}
\begin{equation}
T_{00}^{\text{total}}=\frac{M^{2}}{2\tilde{\kappa}^{2}}\tanh^{2}\bar{x}
\label{21}%
\end{equation}

which, although is finite, and minimizes in the wall's core, gives rise to a
linearly divergent surface energy
\begin{equation}
\mu(|\bar{x}_{c}|)=\frac{1}{M}\int_{-|\bar{x}_{c}|}^{+|\bar{x}_{c}|}%
T_{00}^{\text{total}}d\bar{x}=\frac{M}{\tilde{\kappa}^{2}}|\bar{x}_{c}|
\label{22}%
\end{equation}

where $|\bar{x}_{c}|$ is a large distance cut off.

\section{Summary}

\ \ We have considered the issue of stability for the Gibbons-Wells magnetic
domain wall, which is unusual in that the potential for the scalar field
$V(\phi)=-\frac{1}{2}\mathcal{H}^{2}e^{2\tilde{\kappa}\phi}$, with
$\mathcal{H}=$ const., is a monotonic function of $\phi$, and a normal vacuum
manifold is not identified. This is in contrast to the familiar type of scalar
potential with $V\sim\lambda(\phi^{2}-v^{2})^{2}$ where vacuum states at
$\phi=\pm v$ are identified and there is a $Z_{2}$ discrete symmetry with the
domain wall interpolating between the two vacuum states. In that case, domain
walls are stabilized by topology, whereas the Gibbons-Wells walls are
nontopological. It is therefore reasonable to ask whether these magnetic walls
are stable against small fluctuations. A linear perturbation analysis leads to
the result that a small fluctuation $\delta\phi$ will be static ($\omega=0$)
and localized, and therefore the ansatz solutions $\Phi(x)$ and $\mathbf{B}%
(x)$ for the Gibbons-Wells wall are stable against decay or uncontrolled
growth. Furthermore, the symmetry of $\Phi(x)$ and $\mathbf{B}(x)$, and
therefore the symmetry of the energy density $T_{00}(x)$, indicates a
stability against bending, since there is no biasing in the energy density.

\end{document}